\documentclass[prb,twocolumn,showpacs,amsmath,amssymb,floatfix]{revtex4}
\def\be{\begin{equation}}
\def\ee{\end{equation}}
\def\ber#1\eer{\begin{align}#1\end{align}}
\def\up{\uparrow}
\def\down{\downarrow}
\def\Ev{\mathbf{E}}
\def\nablabold{\mbox{\boldmath $\nabla$}}

\usepackage[dvips]{graphicx} 

\begin{document}
\bibliographystyle{apsrev}
\title{Ambipolar spin diffusion and {D'yakonov-Perel'} spin relaxation in GaAs quantum wells}
\author{Hui Zhao}
\affiliation{Department of Physics and Astronomy, The University of Kansas, Lawrence, Kansas 66045, USA}
\author{Matt Mower and G. Vignale}
\affiliation{Department of Physics and Astronomy, University of Missouri, Columbia, Missouri 65211, USA}

\begin{abstract}
We report theoretical and experimental studies of ambipolar spin diffusion in a semiconductor. A circularly polarized laser pulse is used to excite spin-polarized carriers in a GaAs multiple quantum well sample at 80~K. Diffusion of electron and spin densities is simultaneously measured using a spatially and temporally resolved pump-probe technique. Two regimes of diffusion for spin-polarized electrons are observed. Initially, the rate of spin diffusion is similar to that of density diffusion and is controlled by the ambipolar diffusion coefficient. At later times, the spin diffusion slows down considerably relative to the density diffusion and appears to be controlled by a non-constant (decreasing) spin diffusion coefficient. We suggest that the long-time behavior of the spin density can be understood in terms of an inhomogeneous spin relaxation rate, which grows with decreasing density. The behavior of the spin relaxation rate is consistent with a model of D'yakonov-Perel' relaxation limited by the Coulomb scattering between carriers.
\end{abstract}

\pacs{72.25.Dc,~78.47.jc,~72.25.Fe,~72.25.Rb,~73.63.Hs}

\maketitle

\section{Introduction}

Recently, concerted efforts have been directed at developing new ways of understanding the spin-degree of freedom in semiconductors and its potential use in electronic devices.\cite{s2941488,rmp76323,apl56665} Although electric spin injection has been achieved in a number of semiconductors, such as Si,\cite{n447295} GaAs,\cite{l83203,nphys3197} InAs,\cite{l88066806,apl90022101} carbon nanotubes,\cite{n401572,nphys199} and graphene,\cite{n448571} spin transport in semiconductors is more conveniently studied by optical techniques. Optical spin injection and detection are very efficient due to well-established spin selection rules.\cite{opticalorientation} The transport of spin can be studied directly with high temporal and spatial resolution. For example, spin transport in GaAs has been studied by measuring polarization properties of photoluminescence.\cite{apl731580,apl812788,l87276601,nmat4585,l98036603} Faraday and Kerr rotation techniques have been used to study transport of spins injected by optical orientation,\cite{n397139,n411770,l94236601} by the spin Hall effect,\cite{s3061910,nphys131,nphys4843} and from ferromagnets.\cite{s3092191} Very recently, the spin-grating technique has been used to study pure spin diffusion.\cite{n4371330,l97136602,l98076604,l764793,oc174291} Spin transport in semiconductors has also been an active topic of theoretical studies in recent years.\cite{l844220,b66235302,b624853,b68045307,b74121303,b71125103,b73075306,b74195330,b74155207,b66235109,b64024426}

In optical excitations, the injection of spin-polarized electrons is inevitably accompanied by the injection of an equal number of positively charged and spin-polarized holes.\cite{l844220} In previous experimental studies, however, the effect of the holes was minimized or eliminated by various strategies. In studies of {\it n}-type doped semiconductors, for instance, the optically injected holes quickly lose their polarization and recombine with majority carriers, leaving behind a purely electronic spin packet.\cite{n397139,n411770,l94236601,s3061910,nphys131,nphys4843,s3092191} In intrinsic samples with an externally applied electric field, holes are spatially separated from the electrons, and the transport is dominated by drift of spin-polarized electrons in the field.\cite{apl731580,apl812788} In spin-grating experiments, interference of two oppositely polarized laser beams results in polarization modulation -- without intensity modulation, across the laser spot -- which is typically much bigger than the grating period. Since the excited carrier density is uniform, no carrier diffusion occurs. Spin diffusion in this configuration is not influenced by holes.\cite{n4371330,l97136602,l98076604,l764793,oc174291} Finally, in spin-transport experiments using surface acoustic waves, spatially modulated piezoelectric fields of surface acoustic waves separate and move spin-polarized carriers. Since electrons and holes are trapped at the minima and maxima of the acoustic wave, respectively, holes are separated from electrons over macroscopic distances.\cite{l87276601,nmat4585,l98036603} In all these cases, since the influence of the holes is suppressed, spin transport is dominated by the drift and diffusion of spin-polarized electrons, analog to unipolar carrier transport.

\begin{figure}
\includegraphics[width=7.5cm]{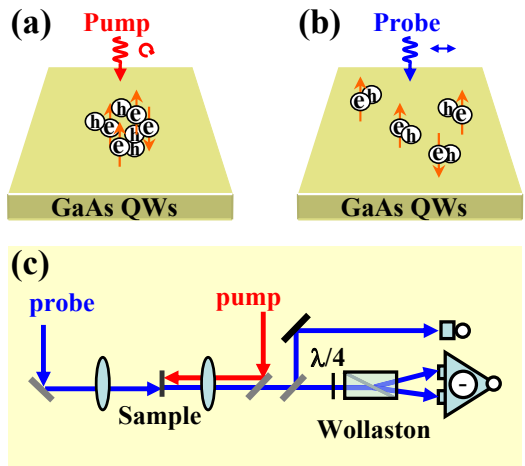}
\caption{(Color online) Experimental geometry (a) Spin-polarized electrons ($e$) and holes ($h$) are injected in GaAs QWs by a circularly polarized pump pulse. Right after the hole spin relaxation, the carrier system is composed of spin-up electrons, spin-down electrons, and holes, within the excitation spot. The arrows on the electrons indicate the direction of the electron spin. This is the initial state of the ambipolar carrier and spin diffusion process to be studied. (b) The carrier system at a later time. Diffusion of electron-hole pairs is driven by the density gradient, and is temporally and spatially resolved by measuring the differential transmission of a linearly polarized probe pulse (c) Experimental set-up. The pump and probe pulses are counter propagating and are focused to the sample by microscope objectives with numerical aperture of 0.4. A portion of the transmitted probe pulse is reflected by a beam splitter to a silicon photodiode to detect electron density. The spin density is detected by analyzing the polarization state of the other portion of the transmitted probe pulse by an optical bridge composed of a quarter-wave plate ($\lambda/4$), a Wollaston prism, and a balanced detector.}
\end{figure}

In this paper, we report theoretical and experimental studies of diffusion of optically injected spin-polarized carriers in an intrinsic semiconductor without an externally applied electric field. In this case, the spin diffusion is strongly influenced by the presence of the holes. In the experiments, spin-polarized electron-hole pairs are excited in GaAs quantum wells (QWs) by a tightly focused and circularly polarized laser pulse. Since hole spin relaxation is shorter than a few picoseconds,\cite{l89146601} after a short transient process, the carrier system has three components, spin-up electrons, spin-down electrons, and holes, as illustrated in Fig.~1a. All three species diffuse in the quantum-well plane and interact with each other. Figure 1b illustrates the carrier system at a later time. Although holes are unpolarized, they influence the diffusion of spin-polarized electrons via the Coulomb attraction. Basically, the rate of diffusion is controlled by the slow holes, while the highly mobile electrons quickly adjust to neutralize the holes. This forces both spin-up and spin-down electrons to diffuse in the same direction, in marked opposition to unipolar spin packets in which electrons of opposite spin orientations diffuse in opposite directions.

We study this triple-polar diffusion process by a pump-probe technique. Dynamics of electron density and spin density are spatially and temporally resolved by measuring carrier-induced changes in the transmission of a linearly polarized probe pulse [Fig.~1b]. We observe a sub-linear-expansion process of the area of the spin density packet, while the simultaneously measured electron density packet expands linearly. This indicates that the spin transport cannot be described as a classical diffusion process with a constant diffusion coefficient. The spin diffusion is significantly slower than the ambipolar carrier diffusion i.e., diffusion of holes screened by electrons. Our theoretical analysis, based on a three-component drift-diffusion equation, shows that the long-time behavior of the spin density can be understood in terms of a spin relaxation rate that grows with decreasing density. This behavior is consistent with a model of D'yakonov-Perel' relaxation limited by the Coulomb scattering between the carriers. However, the short-time behavior of the diffusion, where the spin diffusion coefficient decreases significantly with time, remains for the time being beyond the reach of our drift-diffusion theory, and will be the subject of further investigation.

This paper is organized as follows. In Sec. II, we discuss experimental techniques used for optical injection and detection of spin-polarized carriers. Experimental results are presented in Sec. III. A drift-diffusion theory is outlined in Sec. IV, followed by theoretical and experimental results on an inhomogeneous spin relaxation time in Sec. V. We conclude with a summary in Sec. VI. More details on spin diffusion matrix are discussed in the Appendix.

\section{Experimental techniques}

The GaAs multiple-quantum-well sample contains 40 periods of 10-nm GaAs layers sandwiched by 10-nm $\mathrm{Al_{0.7}Ga_{0.3}As}$ barriers. During the measurements, the sample is cooled to 80~K. Spin-polarized carriers are excited by a circularly polarized 250-fs pump pulse with a central wavelength of 1550~nm obtained from an optical parametric oscillator pumped by a Ti:sapphire laser at 80~MHz. The circular polarization is achieved by a broadband quarter-wave plate and a linear polarizer, with a purity better than 97\% on the sample. The pump pulse is tightly focused to excite carriers with a Gaussian spatial profile of $w_0$=1.5~$\mu$m (full width at half maxima) through two-photon absorption, with an excitation excess energy of about 40~meV. The density of spin-up electrons excited, $n_{\uparrow}$, is approximately three times higher than the density of spin-down electrons, $n_{\downarrow}$, owing to selection rules for two-photon absorption of circularly polarized light.\cite{b71035209} The advantage of using two-photon absorption process instead of one-photon absorption to excite carriers is that, due to a small two-photon absorption coefficient, the peak electron density is uniform across the quantum wells since the intensity of the pump pulse does not attenuate significantly as it propagates through the sample.

The carrier and spin dynamics are monitored by simultaneously measuring electron and spin density profiles as functions of time by using a spatially and temporally resolved pump-probe technique. The total electron density, $n \equiv n_{\uparrow}+n_{\downarrow}$, is measured by focusing a linearly polarized 100-fs probe pulse on the sample with a spot size of 1.5~$\mu$m [Fig.~1c]. The probe pulse is obtained from the Ti:sapphire laser that is used to pump the optical parametric oscillator, and is tuned to $1s$ heavy-hole excitonic resonance (803~nm) to efficiently probe carriers based on well-established excitonic absorption saturation caused by free carriers.\cite{b326601} The differential transmission, $\Delta T/T_0 \equiv [T(n)-T_0]/T_0$, i.e., the normalized difference between transmission in the presence of carriers [$T(n)$] and without them [$T_0$], is measured by reflecting a portion of the transmitted probe pulse to a silicon photodiode [see Fig.~1c]. This quantity is directly proportional to the electron density under our experimental conditions (see discussion below). A lock-in amplifier is used to detect the signal with the pump intensity modulated at 821~Hz by a mechanical chopper.

Spin density, $s \equiv n_{\uparrow}-n_{\downarrow}$, is measured by analyzing carrier-induced circular dichroism of the same probe pulse, i.e., the absorption difference of right- and left-hand circularly polarized probe pulses in the presence of spin-polarized carriers. The linearly-polarized probe pulse is composed of two circular components. Due to spin-selection rules, each component preferentially senses one spin system.\cite{opticalorientation} By using a quarter-wave plate ($\lambda /4$), the two circular components are converted to two orthogonal linear polarizations, which are then separated by a Wollaston prism and sent to two photodiodes of a balanced detector. The output of the balanced detector is proportional to the difference between the differential transmissions of the two circular components, $(\Delta T^+ -\Delta T^-)/T_0$, which is proportional to the spin density $s$.\cite{b72201302,l96246601,b75075305,b78045314}

In order to deduce the actual electron and spin densities from the measured differential transmission signals $\Delta T/T_0$ and $(\Delta T^+ -\Delta T^-)/T_0$, separate measurements are performed with the same set-up, but with a pump pulse of a central wavelength at 775~nm, which is obtained by the second-harmonic generation of the 1550-nm pulse. This pump pulse excites carriers via one-photon absorption. The linear relation between $\Delta T/T_0$ and $n$ is verified by measuring $\Delta T/T_0$ as a function of the energy fluence of this pump pulse, which is proportional to the electron density in the linear regime. Furthermore, we measure the absorption coefficient of the sample at this pump wavelength by comparing the incident, reflected, and transmitted powers of the pump pulse. This allows us to determine the number of photons absorbed by the sample for a given pump energy fluence. Since other absorption processes are much weaker than the interband absorption, approximately the same number of carriers is excited. Therefore, we obtain the proportionality constant between $\Delta T/T_0$ and $n$. The constant of proportionality between the spin density $s$ and $(\Delta T^+ -\Delta T^-)/T_0$ is similarly deduced, using in addition the well-established fact that the one-photon absorption of circularly polarized light tuned close to the band gap of GaAs excites electrons with a spin polarization of $s/n=0.5$.\cite{opticalorientation}

\section{Experimental results}

With the experimental techniques discussed in Sec. II, we simultaneously measure electron density and spin density as functions of space by scanning the probe spot across the pump spot, and as functions of time by scanning the time-delay between the probe and the pump pulses. Figure 2 summarizes the dynamics of electron density and spin density measured with a peak electron density of $2.3 \times \mathrm{10^{17}~cm^{-3}}$. The contour map in Fig.~2a shows how electron density varies with time and space after injection. Here $r$ is defined as the distance between the centers of the pump and probe spots, and $t$ is the time delay of the probe pulse with respect to the pump pulse. Two normalized spatial profiles of electron density measured at  $t$=5~ps (squares) and 300~ps (circles) are plotted in Fig.~2b. The solid lines are Gaussian fits to the data. Clearly, the spatial profile of electron density remains Gaussian shape, and expands due to the transport. To quantitatively describe the transport, we deduce the time variation in the full width at half maxima of the electron density profile, $w_{\mathrm{n}}$, by fitting the electron density profiles measured at all probe delays. In Fig~2c, the squared width as a function of $t$ is plotted.

\begin{figure}
\includegraphics[width=8.5cm]{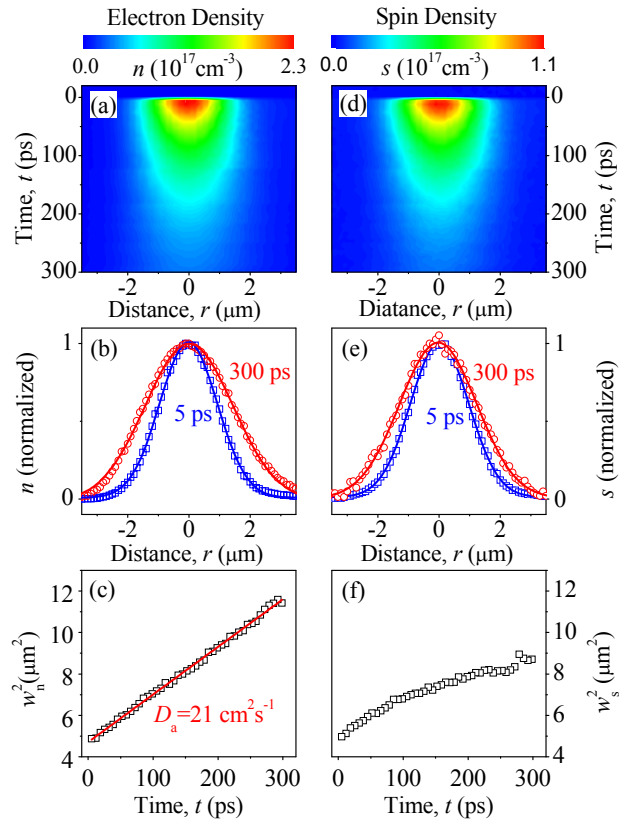}
\caption{(Color online) Spatiotemporal dynamics of electron ($n$) and spin ($n$) densities measured at a sample temperature of 80~K and a peak electron density of $2.3 \times \mathrm{10^{17}~cm^{-3}}$. (a) Electron density as functions of time and space. The electron density is deduced from the measured $\Delta T/T_0$. (b) Spatial profiles of electron density measured at 5 ps (squares) and 300 ps (circles). The profiles are normalized in order to show the expansion caused by the carrier diffusion. The solid lines are Gaussian fits. (c) Squared width of electron density profiles as a function of time obtained by Gaussian fit to profiles measured at various times. The linear fit (solid line) corresponds to an ambipolar carrier diffusion coefficient of $D_a=21~\mathrm{cm^2 s^{-1}}$. (d)-(f) Spatiotemporal dynamics of spin density obtained in the same scans.}
\end{figure}

According to the classical diffusion model with a constant diffusion coefficient, the squared width increases linearly as
\begin{equation}
\label{diffusiveexpansion}
	w^{2}_{\mathrm{n}}(t)=w^{2}_{\mathrm{0}}+16\ln (2)D_at,
\end{equation}
where $D_a$ is the ambipolar carrier diffusion coefficient.\cite{b385788,b67035306} By linear fit to the data, we deduce $D_a=21~\mathrm{cm}^2 \mathrm{s}^{-1}$. It is worth mentioning that, in the experiment, due to a finite size of the probe spot, the measured profiles shown in Fig.~2b are actually convolutions of the probe spot and the actual electron density profiles. However, since both the probe spot and the electron density profiles are Gaussian, the convolution does not influence the measurement of $D_a$.\cite{apl92112104}

The right half of Fig.~2 shows the spin component of the diffusion process measured simultaneously with density. The contour map in Fig.~2d shows the spatiotemporal dynamics of spin density after optical excitation. Spin diffusion is evident by comparing the two profiles in Fig.~2e measured at $t$=5~ps (squares) and 300~ps (circles). The profile remains Gaussian, as confirmed by the fits (solid lines). However, the spin density profile at $t$=300~ps is narrower than the electron density profile shown in Fig.~2b. This indicates the different transport behavior of spin and electron densities. Quantitatively, the squared width of the spin density profile does not increase linearly, as shown in Fig.~2f, in striking contrast to the expansion of the electron density profile obtained in the same measurement. Initially, the expansion rate of spin density profile is similar to that of electron density profile. At later times, the spin diffusion slows down considerably relative to the density diffusion. The observed sub-linear expansion of spin density profile shows that spin diffusion cannot be described as a classical diffusion process with a {\it constant} diffusion coefficient.

\begin{figure}
\includegraphics[width=6.5cm]{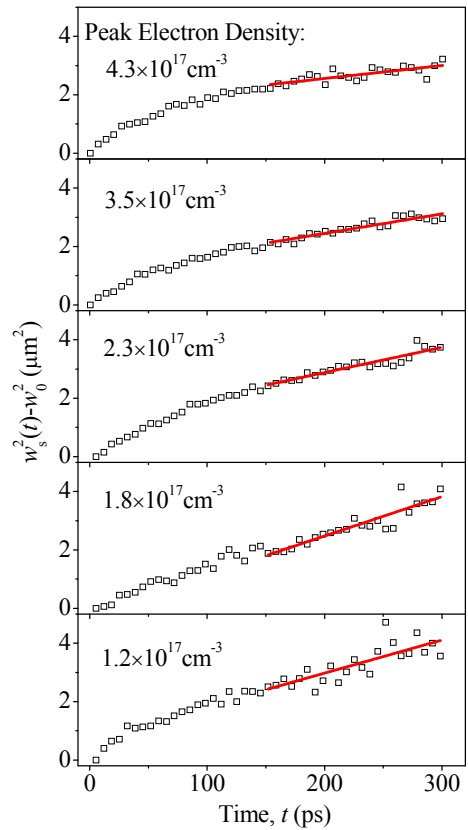}
\caption{(Color online) Expansions of spin density profiles at several peak electron densities as indicated in each panel, obtained by repeating the procedure summarized in Fig.~2. The sublinear expansion is generally observed. Linear fits to data with $t > 150$ ps (solid lines) are used in order to approximately illustrate the density dependence of spin diffusion.}
\end{figure}

The procedure summarized in Fig.~2 is used to systematically study the influence of electron density on the diffusion process. The sub-linear expansion of the spin density profile is observed at all densities and changes systematically with the peak electron density. In Fig.~3 we show several examples. The decrease in the slope with time is more pronounced at higher densities. Meanwhile, the simultaneously obtained expansions of the electron density profile (not shown) are all linear and similar to Fig.~2c. To approximately compare the rates of spin diffusion and carrier diffusion, we perform linear fits to the data with $t > 150$ ps (solid lines in Fig.~3). In this range the expansion is approximately linear. The spin diffusion coefficients deduced by the fits are plotted in Fig.~4 (circles) as a function of peak electron density. The ambipolar carrier diffusion coefficients are also plotted (squares) for comparison. In the density range studied, the ambipolar carrier diffusion coefficient is almost constant. In contrast, the spin diffusion coefficient decreases significantly with peak electron density.

\begin{figure}
\includegraphics[width=8cm]{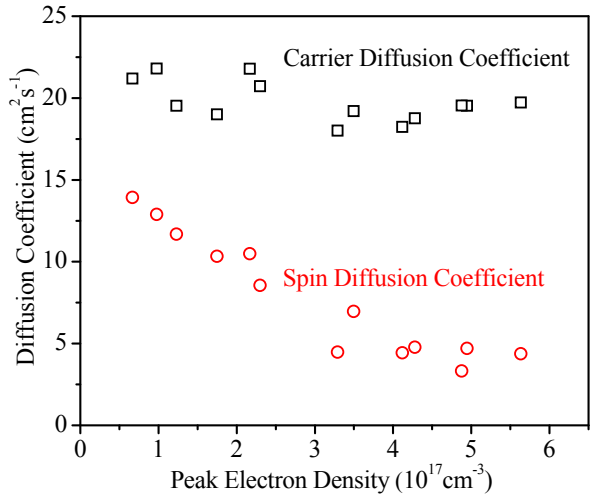}
\caption{(Color online) Ambipolar carrier diffusion coefficient (squares) and spin diffusion coefficient (circles) as functions of peak electron density obtained by repeating the procedure summarized in Fig.~2 at various electron densities. These coefficients are deduced by linear fits as shown in Figs. 2 and 3. Significant decrease in spin diffusion coefficient with peak electron density is observed. The carrier diffusion coefficient is almost constant.}
\end{figure}

\section{Drift-diffusion theory}
We denote by $p$ the density of holes. The drift-diffusion equations for the three coupled densities, $n_{\uparrow}$, $n_{\downarrow}$, and $p$, are\cite{bookmqcsss}
\begin{align}
	\label{DDE_n_up_FULL}
	\frac{\partial n_{\uparrow}}{\partial t} =& \nablabold \cdot \left[\frac{\sigma_\up}{e}\Ev + D_{\uparrow\uparrow} \nablabold n_{\uparrow} + D_{\uparrow\downarrow} \nablabold n_{\downarrow} \right] + \frac{n_{\downarrow}}{\tau_{\downarrow\uparrow}} - \frac{n_{\uparrow}}{\tau_{\uparrow\downarrow}} - \frac{n_{\uparrow}}{\tau_{r}} \\
	\label{DDE_n_down_FULL}
	\frac{\partial n_{\downarrow}}{\partial t} =& \nablabold \cdot \left[\frac{\sigma_\down}{e} \Ev + D_{\downarrow\downarrow} \nablabold n_{\downarrow} + D_{\downarrow\uparrow} \nablabold n_{\uparrow} \right] + \frac{n_{\uparrow}}{\tau_{\uparrow\downarrow}} - \frac{n_{\downarrow}}{\tau_{\downarrow\uparrow}} - \frac{n_{\downarrow}}{\tau_{r}} \\
	\label{DDE_p_FULL}
	\frac{\partial p}{\partial t} =& \nablabold\cdot \left[ -\frac{\sigma_p}{e} \Ev + D_p \nablabold p \right] - \frac{p}{\tau_{r}}
\end{align}
where $\sigma_{\up}$, $\sigma_{\down}$ and $\sigma_p$ are the ordinary conductivities of spin-up electrons, spin-down electrons and holes respectively; $D_{\sigma\sigma'}$ is the spin-diffusion matrix for electrons; $\Ev$ is the electric field; $1/\tau_{\up\down}$ and $1/\tau_{\down\up}$ are the spin flip rates from up to down and from down to up, respectively; $\tau_{r}$ is the electron-hole recombination time.

Some physical assumptions underlie the above equations. First of all, we have completely neglected the spin polarization of the holes, assuming that the spin relaxation time of the holes is very short. Second, we have ignored the influence of carrier thermalization and relaxation on the diffusion process, and assumed a constant carrier temperature. This is justified since these processes are expected to end on the time scale of 10~ps even with the influence of hot-phonon effect.\cite{rpp54169} Indeed, the linear expansion of carrier density profile observed in Fig.~2c confirms that the diffusion coefficient is a constant, and the influence of thermalization and relaxation on diffusion is negligible although the electrons are initially excited with an excess energy of 40~meV. Finally, by keeping the off-diagonal elements of the spin-diffusion matrix (e.g. $D_{\up\down}$) we have allowed in principle for the effect of {\it spin Coulomb drag},\cite{PRB65085109} whereby a gradient of spin-up density can drive a current of spin down and vice-versa. The explicit form of this matrix is
\begin{equation}
	\hat{D} = \frac{D_n}{1+\gamma\tau}
	\begin{bmatrix}
		1+\frac{n_{\up}}{n}\gamma \tau & \frac{n_{\up}}{n}\gamma \tau \\
		\frac{n_{\down}}{n}\gamma \tau & 1+\frac{n_{\down}}{n}\gamma \tau
	\end{bmatrix}
\end{equation}
where $D_n = \mu_n k_BT/e$ is the electron diffusion constant, $\tau$ is the momentum relaxation time for electrons, and $\gamma$ is the spin Coulomb drag coefficient. The steps for deriving the spin diffusion matrix from the spin resistivity matrix are outlined in the Appendix.

The equations for $n_{\up}$ and $n_{\down}$ can be rewritten in terms of total density $n \equiv n_{\uparrow}+n_{\downarrow}$ and spin density $s \equiv n_{\uparrow}-n_{\downarrow}$ as follows:
\begin{align}
	\label{SPE_ASYM_n_dd}
	\frac{\partial n}{\partial t} =& \nablabold \cdot \left[\mu_n n \Ev + D_n \nablabold n \right] -  \frac{n}{\tau_{r}} \\
	\label{SPE_ASYM_s_dd_ud}
	\frac{\partial s}{\partial t} =& \nablabold \cdot \left[\mu_n s \Ev + D_s \left( \nablabold s + s \gamma\tau \nablabold \ln n \right) \right] - \frac{s}{\tau_s} - \frac{s}{\tau_{r}}\,,
\end{align}
where $\mu_n$ and $\mu_p$ are the mobilities of electrons and holes, respectively, and we have used the standard relations $\sigma_{\up(\down)} = n_{\up(\down)} e \mu_n$ and $\sigma_p = ep\mu_p$. In the second equation we have set
\begin{equation}
	\label{spindiffusionconstant}
	D_s = \frac{D_n}{1+\gamma\tau}
\end{equation}
and have assumed $1/\tau_{\up\down}=1/\tau_{\down\up}$. Finally, $1/\tau_{\up\down}+1/\tau_{\down\up}=1/\tau_s$, where $1/\tau_s$ is the spin relaxation rate.

The electric field $\Ev$ arises from the small charge imbalance that inevitably occurs as low-mobility holes try to keep up with high-mobility electrons. Even though the charge imbalance is very small, it would not be legitimate to neglect this field, since the conductivity of the electrons is very high. Comparing the equations for $n$ [\ref{SPE_ASYM_n_dd}] and $p$ [\ref{DDE_p_FULL}], and making use of the approximate charge neutrality condition $p \simeq n$, we identify the electric field as
\begin{equation}
	\label{EFA_E_approx}
	\Ev = \frac{D_p-D_n}{\mu_p+\mu_n}\frac{\nablabold n}{n}\,.
\end{equation}
Given the relative magnitudes of $D_p < D_n$, we can see that the electric field points away from high density regions. The impact of this small electric field is a slight resistance to the diffusion of electrons.

Substituting this into Eqs.~(\ref{SPE_ASYM_n_dd}) and~(\ref{SPE_ASYM_s_dd_ud}) we arrive at our final equations:
\begin{align}
	\label{SPE_ASYM_n_dd2}
	\frac{\partial n}{\partial t} =& D_a \nabla^2 n - \frac{n}{\tau_{r}}\\
	\label{SPE_ASYM_s_dd_ud2}
	\frac{\partial s}{\partial t} =& \nablabold \cdot \left[(D_a-D_s) s \nablabold \ln n + D_s \nablabold s \right] - \frac{s}{\tau_s} - \frac{s}{\tau_{r}}\,,
\end{align}
where
\begin{equation}
	D_a = \frac{D_n\mu_p + D_p\mu_n}{\mu_n + \mu_p}
\end{equation}
is the {\it ambipolar} diffusion constant, which is intermediate between $D_n$ and $D_p$ but numerically closer to the diffusion constant of the less mobile species (holes in this case). Equation (\ref{SPE_ASYM_n_dd2}) is the standard electron diffusion equation with ambipolar diffusion constant. Strictly speaking, the recombination time can vary with density, complicating the solution. Experimentally, a recombination time between 160 ps at low densities ($\sim \mathrm{10^{17}~cm^{-3}}$) and 190 ps at higher densities ($\sim 6 \times \mathrm{10^{17}~cm^{-3}}$) was observed. For the time being, consider the recombination time to be spatially uniform, a point to which we will return and justify later. Assuming that the electron density packet at the initial time $t=0$ has a Gaussian shape, $n(r,0) = N e^{-4\ln (2)r^2/w_0^2}$, we see that the solution to Eq.~(\ref{SPE_ASYM_n_dd2}) at time $t$ is given by
\begin{align}
	n(r,t)=N \left(\frac{w_0^2}{w_{\mathrm{n}}^2(t)}\right)^{3/2} e^{ -4\ln (2)r^2/w_{\mathrm{n}}^2(t)- t / \tau_{r}} ,
\end{align}
where $w_{\mathrm{n}}^2(t)=w_0^2+16\ln (2)D_at$. The linear growth in time of the area covered by the packet is in excellent agreement with the experimental observation, as shown in Fig.~2c.

Let us now consider Eq.~(\ref{SPE_ASYM_s_dd_ud2}) for the spin density. We assume that the initial spin distribution is proportional to the density distribution, i.e., we have
\begin{equation}\label{initialcondition1}
	s(r,0)=Cn(r,0)
\end{equation}
where $C$ is a constant independent of position. For the present experiment the expected value of $C$ is $C=1/2$. The number of spin-up electrons is three times larger than the number of spin-down electrons, owing to selection rules for two-photon absorption of circularly polarized light.\cite{b71035209} Then, {\it if the spin relaxation time is neglected} we see immediately -- by direct substitution -- that the solution of Eq.~(\ref{SPE_ASYM_s_dd_ud2}) is
\begin{equation}
	\label{solutiontrivial}
	s(r,t)=Cn(r,t)\,.
\end{equation}
In other words, the spin density and the ordinary density diffuse at exactly the same rate, controlled by $D_a$.

It should be noted that this result holds irrespective of the value of the spin Coulomb drag coefficient. The ambipolar spin diffusion in our configuration is in sharp contrast with the case of {\it unipolar} spin packets in spin-grating experiments.\cite{n4371330,l97136602,l98076604,l764793,oc174291} In those experiments, two laser pulses that are oppositely linearly polarized are used to excited carriers. Since the interference does not introduce any intensity modulation of the light, the total electron density and hole density excited are not modulated. Hence, there is no density gradient to cause diffusion of holes and electrons. However, the interference gives rise to spatial modulations of the light {\it polarization}, resulting in periodic variations in spin density of electrons. The excited spin grating decays due to the diffusion of spin-up and spin-down electrons from high density to low density regions, within the laser spot. In this case, the same theory predicts the spin diffusion constant to be $D_s$, given by Eq.~(\ref{spindiffusionconstant}).\cite{n4371330} In this {\it unipolar} spin diffusion regime, the requirement of charge neutrality forces the electrons of opposite spin orientations to move in opposite directions: spin Coulomb drag arises from this relative motion. In the ambipolar spin- diffusion process, the spin-density profile overlaps with the hole density profile, and charge neutrality is ensured by the holes. Thus, the electrons move in the same direction regardless of spin - there is no relative motion and therefore no spin Coulomb drag, and the diffusion is controlled by the ambipolar diffusion coefficient.

The above solution~(\ref{solutiontrivial}) was found for the initial condition~(\ref{initialcondition1}) but we have checked that, even if the spin density is not initially proportional to the density, it eventually becomes proportional to the density at sufficiently long times.

We conclude that the experimental observation of an apparently decreasing spin diffusion coefficient cannot be explained within the framework of the drift-diffusion theory, unless one is willing to include the spin relaxation time. Further, it is easy to see that a {\it homogeneous} spin relaxation time ($1/\tau_s$ independent of density and hence of position) will not change the situation, for the solution of Eq.~(\ref{SPE_ASYM_s_dd_ud2}) in the presence of such a relaxation time is simply
\begin{equation}
	\label{solutiontrivial2}
	s(r,t)=C n(r,t)e^{-t/\tau_s}\,.
\end{equation}
There is now decay of the spin due to spin relaxation in addition to the global decay of carrier and spin densities due to electron-hole recombination.

\section{Inhomogeneous spin relaxation}

In view of the above discussion, we now examine the possibility of explaining the experimental data in terms of non-homogeneous spin relaxation. Suppose, for instance, that the spin relaxation rate were larger at lower density, i.e., larger in the tails of the spin packet than at its center. This would produce the impression of slower spin diffusion, since the outward diffusion of the spin would be hidden by the more rapid decay of the spin at the edges (Fig.~5). To verify this idea, we measure the spin relaxation time as a function of peak electron density by using the same pump-probe technique. The results shown in Fig.~7 confirm this conjecture (see discussion below). The spin relaxation time is larger for packets of higher density, suggesting that $1/\tau_s$ increases with decreasing density. However, these are measurements of the lifetime of the spin packet as a whole, whereas in the drift-diffusion theory we need a position-dependent spin relaxation rate, determined by the local density.

\begin{figure}
\includegraphics[width=8cm]{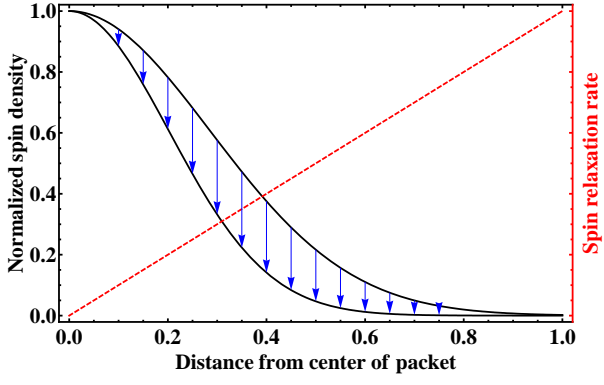}
\caption{(Color online) As electron density decreases, the spin relaxation rate increases (dashed line). A non-uniform spin relaxation rate across the packet gives the appearance of slower width increase. The length of the arrows is given by the product of density and relaxation rate. (linear $1/\tau_s$ for illustrative purposes only)}
\end{figure}

We have developed a model for the position dependence of the spin relaxation rate along the following lines. First, we notice that electron-impurity effects or spin-orbit interactions with the lattice could not account for the spatial variation of $1/\tau_s$ within the packet, since the impurity environment and the crystalline environment are uniform over the region occupied by the packet. What is not uniform is the frequency of electron-electron (e-e) and electron-hole (e-h) collisions. Because the Coulomb interaction is screened and thus effectively becomes of short range, we can say that the carriers near the center of the packet experience an electronic environment of higher density, and their scattering rate is accordingly higher than in the tails of the packet, where they experience an electronic environment of lower density.

How does the variation in the carrier-carrier scattering time translate into the observed spatial variation in the spin relaxation rate? A plausible answer comes from the D'yakonov-Perel' spin relaxation mechanism.\cite{opticalorientation,jsinm16735} In this mechanism, the spin relaxation is primarily due to the spin-orbit interaction with the lattice. In GaAs the primary mechanism is the independent precession of each electron in the Dresselhaus $k$-dependent effective magnetic field\cite{pr100580}
\begin{align}
	\label{Dresselhaus}
	\mathbf{\Omega_k} =& \alpha_c \hbar^2 \left(2 {m^*}^3 E_g\right)^{-1/2} \boldsymbol{\kappa}\,, \\
	\boldsymbol{\kappa} =& \left\{ k_x \left( k_y^2-k_z^2 \right),\, k_y \left( k_z^2-k_x^2 \right),\, k_z \left( k_x^2-k_y^2 \right) \right\}
\end{align}
where the parameter $\alpha_c\simeq 0.07$ for GaAs, $m^*$ is the effective mass of electrons in the conduction band, and $E_g$ is the band-gap energy. Notice that the effective magnetic field depends on the momentum ${\bf k}$ of the electron. e-e and e-h collisions change the momentum of the electron, and thus change the direction and magnitude of ${\bf \Omega_k}$. This hinders the process of spin relaxation, especially in regions where momentum-changing collisions are frequent, i.e. high-density regions. A standard analysis for cubic systems leads to the conclusion that
\begin{equation}
	\label{DPFormula}
	\frac{1}{\tau_s} = \langle \Omega_{\perp k}^2\tau^*\rangle \simeq \frac{32}{105}\tau^* \alpha_c^2 \frac{\epsilon_k^3}{\hbar^2 E_g}
\end{equation}
where $\tau^*$ is related to the momentum relaxation time by the effectiveness of the scattering event on the randomization of the axis of $\Omega_k$, $\Omega_{\perp k}$ is the component of ${\bf \Omega}_k$ perpendicular to the spin axis, $\epsilon_k=\hbar^2k^2/2m^*$ is the one-particle energy, and the angular bracket denotes a thermal average in momentum space.

\begin{figure}
\includegraphics[width=8cm]{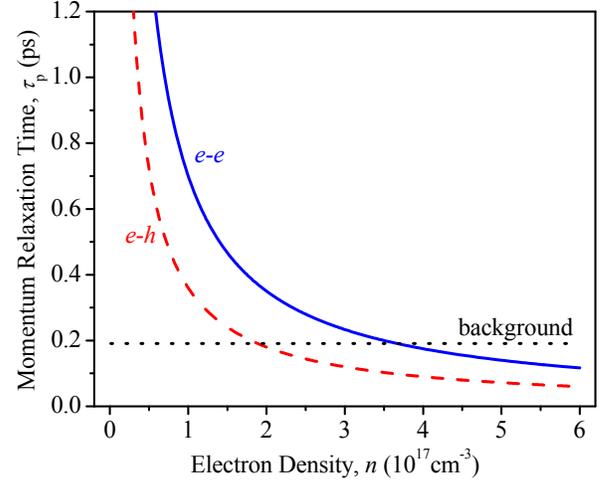}
\caption{(Color online) The momentum relaxation times due to e-e (solid line) and e-h (dashed line) scatterings are on the same order of magnitude as the background scattering (dotted line) in high density regions ($>10^{17}$cm$^{-3}$) but become quite long at lower densities. Under the present interaction model, the rate of carrier-carrier collisions is proportional to $n$. Here the relaxation time is calculated for wave vector $k=1.43 \times 10^8$ m$^{-1}$ and screening wave vector $k_0 = 1.89 \times 10^8$m$^{-1}$.}
\end{figure}

We have evaluated the contribution of e-e and e-h interactions to the lifetime of a momentum state in a homogeneous environment of density $n$ according to standard formulas reported for example in Ref.~\cite{bookqtel}, but evaluated at the parameters of the experiment. In order to simplify evaluation of the momentum relaxation time we use a screened Coulomb interaction model of the form
\begin{equation*}
\frac{e^2}{\epsilon_r r}e^{-k_0 r}
\end{equation*}
where $k_0$ is a fitting parameter of the order of the Debye screening wave vector at $n=\mathrm{10^{17}~cm^{-3}}$ and Boltzmann statistics are used throughout. This is admittedly a significant approximation as the electrons are degenerate above $\sim 10^{17}$cm$^{-3}$. While numerical accuracy may be improved, qualitative aspects are generalizable. Our results are shown in Fig.~6, where the momentum relaxation times due to e-e and e-h scatterings are compared to the background scattering due to impurities, which is density independent. These values have been obtained for a typical thermal value of $k= 1.43 \times 10^8$ m$^{-1}$ with a screening wave vector of $k_0 = 1.89 \times 10^8$ m$^{-1}$. Applying Matthiessen's rule, high density regions exhibit a higher rate of collisions which could potentially slow spin relaxation, whereas low density regions are governed primarily by lattice scattering.

\begin{figure}
\includegraphics[width=8cm]{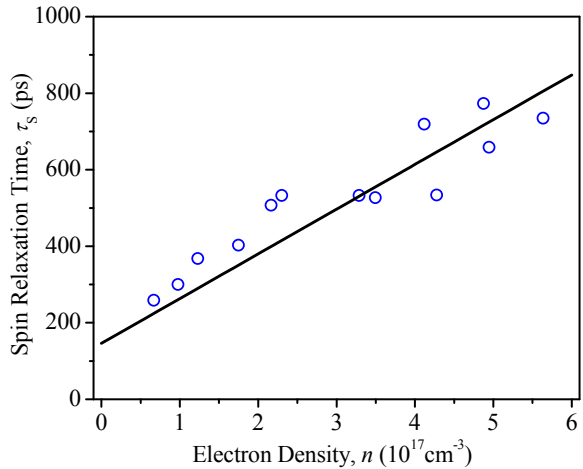}
\caption{(Color online) The spin relaxation time of electrons is inversely proportional to the momentum relaxation time, thus increases with density. Experimentally measured $\tau_s$ (circles) are plotted versus the peak electron density.}
\end{figure}

Using Eq.~(\ref{DPFormula}), we arrive at $\tau_s$ shown in Fig.~7, where the calculated spin relaxation time (solid line) is shown to be in good agreement with experimentally obtained $\tau_s$ (circles). We remind the reader that experimental $\tau_s$ is plotted as a function of peak electron density, while Eq.~(\ref{DPFormula}) is being applied on a microscopic scale. At this point, the earlier approximation of a spatially uniform $\tau_{r}$ can be justified. Whereas the range in spin relaxation time for $\mathrm{10^{17}~cm^{-3}} < n < 6 \times \mathrm{10^{17}~cm^{-3}}$ varies by as much as 600 ps, the electron-hole recombination time only varies by 30 ps. Clearly, the spatial dependence of the recombination rate is swamped off by the spatial dependence of the spin relaxation.

The results of the calculation give us confidence that the D'yakonov-Perel' mechanism, limited by the carrier-carrier scattering, can indeed produce a spin relaxation of the right order of magnitude and -- more importantly for us -- one that is faster in the tails of the packet. We note that our results are consistent with a recent study where electron spin relaxation time in bulk GaAs at room temperature was found to increase linearly with carrier density when the D'yakonov-Perel' mechanism dominates.\cite{apl93132112}

\begin{figure}
\includegraphics[width=8cm]{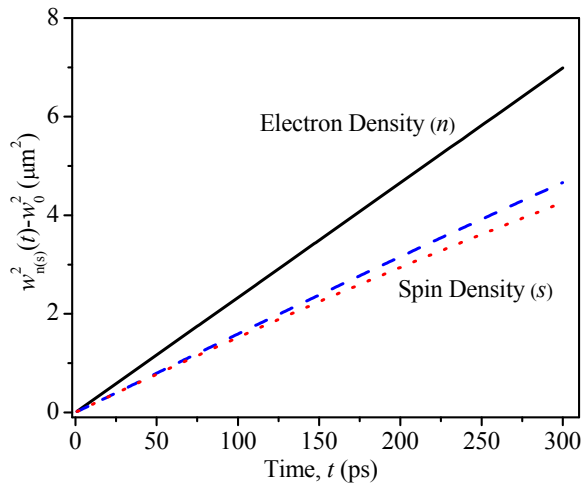}
\caption{(Color online) Squared width of the spin density packet calculated with (dotted line) and without (dashed line) taking into consideration the spin Coulomb drag. These are considerably smaller than the squared width of a density packet (solid line). It is evident that the spin Coulomb drag plays only a minor role in further reducing the apparent diffusion rate. Parameters used matched those of the experiment in Fig.~2.}
\end{figure}

With the inhomogeneous spin relaxation time, the spatiotemporal evolution of a spin polarized packet was evaluated using Eq.~(\ref{SPE_ASYM_s_dd_ud2}) with parameters similar to those used in Fig.~2. One dimensional dynamics were used for the simple goal of comparing diffusion rates. Figure 8 illustrates the squared widths of the profiles of electron density and spin density over an interval of $300$ ps, calculated by fitting a Gaussian curve at several times during the evolution. An interesting consequence of the inhomogeneous spin relaxation is that by breaking the trivial solution it ``turns on" the spin Coulomb drag, which would otherwise be completely inoperative. The spin diffusion rates with and without spin Coulomb drag are also compared. The differences are rather small, reflecting the fact that the slowing down of the spin is mostly apparent, i.e. due to the loss of spin in the edges rather than to the relative motion between up and down spin components. Nonetheless, a small effect is visible and goes in the direction of further reducing the apparent spin diffusion.

\section{Summary}

We have studied, both theoretically and experimentally, the diffusion of optically injected spin-polarized carriers in undoped GaAs quantum wells at 80~K. The experiment is performed with a high-resolution optical pump-probe technique. Spatiotemporal dynamics of locally injected spin-polarized carriers are directly resolved. By comparing expansions of electron and spin density profiles, we found that the spin diffusion cannot be described as a classical diffusion process with a constant diffusion coefficient. The spin diffusion appears to be slower than the ambipolar carrier diffusion. Our theoretical analysis, based on a three-component drift-diffusion equation, shows that the long-time behavior of the spin density can be understood in terms of a spin-relaxation rate that grows with decreasing density. This behavior is consistent with a model of D'yakonov-Perel' relaxation limited by the Coulomb scattering between carriers. However, the short-time behavior of the time varying spin diffusion coefficient (within 100 ps) remains for the time being beyond the reach of our drift-diffusion theory, and is subject to our further investigation.

\section*{Acknowledgments}
We acknowledge John Prineas of University of Iowa for providing us with high-quality GaAs samples. This work was supported bunder NSF Grant No. DMR-0705460 and General Research Fund of The University of Kansas.

\appendix

\section{Spin diffusion matrix}
The spin diffusion matrix is derived from the resistivity matrix by the application of Einstein's relations. The resistivity matrix is given by\cite{PRB65085109}
\begin{equation}
	\hat{\rho} = \frac{m^*}{n e^2 \tau}
	\begin{bmatrix}
		\frac{n}{n_{\uparrow}}+\frac{n_{\downarrow}}{n_{\uparrow}}\tau\gamma & -\tau\gamma \\
		-\tau\gamma & \frac{n}{n_{\downarrow}}+\frac{n_{\uparrow}}{n_{\downarrow}}\tau\gamma
	\end{bmatrix}
\end{equation}
where we have assumed no external electric field. In addition, the momentum relaxation rate due to electron-impurity collisions which flip the spin was taken as negligible in comparison to non-flip processes. $\tau$ is essentially the Drude scattering time.

Inverting the resistivity to find conductivity ($\hat{\sigma}$), Einstein's relation gives the diffusion matrix
\begin{equation}
	e^2 D_{\alpha \beta} = \sum_{\gamma} \sigma_{\alpha \gamma} \left[ \chi^{-1} \right]_{\gamma \beta}
\end{equation}
where the spin susceptibility matrix $\hat{\chi}$ is approximated as
\begin{equation}
	\hat{\chi} =
	\begin{bmatrix}
		\frac{\partial n_{\uparrow}}{\partial \mu_{c\uparrow}} & \frac{\partial n_{\uparrow}}{\partial \mu_{c\downarrow}} \\
		\frac{\partial n_{\downarrow}}{\partial \mu_{c\uparrow}} & \frac{\partial n_{\downarrow}}{\partial \mu_{c\downarrow}}
	\end{bmatrix}
	\cong
	\begin{bmatrix}
		\frac{n_{\uparrow}}{kT} & 0 \\
		0 & \frac{n_{\downarrow}}{kT}
	\end{bmatrix}\,.
\end{equation}
The diffusion matrix is finally,
\begin{equation}
	\hat{D} = \frac{D_n}{1+\gamma\tau}
	\begin{bmatrix}
		1+\frac{n_{\uparrow}}{n}\tau\gamma & \frac{n_{\uparrow}}{n}\tau\gamma \\
		\frac{n_{\downarrow}}{n}\tau\gamma & 1+\frac{n_{\downarrow}}{n}\tau\gamma
	\end{bmatrix}
\end{equation}
where the following relation was used,
\begin{equation}
	D_n = \mu_n \frac{k_B T}{e} = \frac{e \tau}{m^*} \frac{k_B T}{e}\,.
\end{equation}

\end{document}